\documentclass[9pt,conference]{IEEEtran}
\usepackage[preprint]{dcase2025}

\usepackage{bm} % for bold math symbols (incl. Greek letters) with \bm{}

\usepackage{xcolor}
\definecolor{yonghyun}{RGB}{0, 0, 0}

\usepackage{dcase2025,amsmath,graphicx, times,booktabs, tabularx, paralist}
\usepackage{hyperref} 
\usepackage{microtype} 
\title{Audio‑Based Pedestrian Detection in the Presence of Vehicular Noise}

\name{Yonghyun Kim$^{1}$,
      Chaeyeon Han$^{2}$,
      Akash Sarode$^{3}$,
      Noah Posner$^{2}$,
      Subhrajit Guhathakurta$^{2}$,
      Alexander Lerch$^{1}$}
\address{$^{1}$Music Informatics Group, Georgia Institute of Technology, USA\\
$^{2}$Center for Urban Resilience and Analytics, Georgia Institute of Technology, USA\\
$^{3}$College of Computing, Georgia Institute of Technology, USA
}

\begin{document}

\maketitle

\begin{abstract}
Audio-based pedestrian detection is a challenging task and has, thus far, only been explored in noise-limited environments. We present a new dataset, results, and a detailed analysis of the state-of-the-art in audio-based pedestrian detection in the presence of vehicular noise. In our study, we conduct three analyses: 
\begin{inparaenum}[(i)]
\item cross-dataset evaluation between noisy and noise-limited environments, 
\item an assessment of the impact of noisy data on model performance, highlighting the influence of acoustic context, and
\item an evaluation of the model's predictive robustness on out-of-domain sounds.
\end{inparaenum}
The new dataset is a comprehensive 1321-hour roadside dataset. It incorporates traffic-rich soundscapes. Each recording includes \SI{16}{kHz} audio synchronized with frame-level pedestrian annotations and \SI{1}{fps} video thumbnails.
\end{abstract}


\section{Introduction}
\label{sec:intro}

Pedestrian volume data offer valuable insights into urban activity patterns, which support planning efforts such as evaluating sidewalk improvements, assessing land use changes, and identifying areas needing investments in safety and walkability \cite{apa1965pedestriancount}. These data also support optimizing street connectivity and accessibility \cite{apa1965pedestriancount}.

The widespread adoption of smartphones has brought new opportunities for automated human mobility sensing, particularly through mobile GPS data. However, growing privacy concerns, particularly under frameworks like the General Data Protection Regulation (GDPR) in the European Union, have placed restrictions on using mobile location data to track individuals \cite{gdpr2019guidelines}. In parallel, smart city initiatives have adopted the deployment of IoT-based sensors to monitor activity in urban environments. These efforts have largely focused on vision-based systems, such as computer vision and infrared cameras \cite{li2016pedestrian}, although other sensing technologies have also been tested.

Urban sound offers a promising alternative. Microphones are affordable, energy-efficient, and effective in visually occluded environments. They can complement or replace cameras in contexts where installation is impractical, such as shaded areas, narrow corridors, or locations or scenarios where the costs of cameras are prohibitive. The general feasibility of using microphone recordings for the detection of pedestrians has been shown recently for a vehicle-free courtyard on a university campus \cite{seshadri2024asped}.   

This study addresses two key gaps in existing work. First, the generalizability of audio-based models remains unclear. Given the variability in urban soundscapes, shaped by traffic, land use, and average pedestrian activity levels, it is necessary to evaluate model performance across data collected from different settings, particularly in the presence of typical urban noise. Second, existing studies lack information on interpretability; it is unclear which sound characteristics existing models rely on for detecting pedestrians.

Thus, the main contributions of this study are
\begin{compactenum}[(i)]
\item a new publicly available\footnote{\url{https://huggingface.co/datasets/urbanaudiosensing/ASPEDvb}} dataset for audio-based pedestrian detection in the presence of vehicular noise,
\item an investigation into how vehicular noise affects pedestrian detection performance, and
\item insights into the acoustic features that enable pedestrian detection.
\end{compactenum}

\section{RELATED WORK}
\label{sec:related_work}
\subsection{Automated Pedestrian Detection Techniques}
Urban pedestrian sensing technologies have evolved over decades, with video cameras and infrared sensors being the most widely deployed to date \cite{yang2010automatic, li2016pedestrian, yang2011infrared}. Video-based systems, now commonly augmented with computer vision and deep learning techniques, offer high spatial precision but can suffer from limitations in occluded or low-light environments. Furthermore, such systems often raise privacy concerns \cite{BRUNETTI201817, 5975165}. Infrared counters, including active, passive, and target-reflective types, are less intrusive but tend to undercount pedestrians, particularly in high pedestrian volume scenarios \cite{yang2010automatic, yang2011infrared}. More sophisticated but cost-prohibitive options, such as radar, piezoelectric strips, and inductive loops, are limited in spatial scalability \cite{ozan2021}. In contrast, audio-based pedestrian sensing remains underexplored, albeit with promising low-cost deployment, resilience to visual obstructions, and potential privacy advantages. As demonstrated by Seshadri et al.~\cite{seshadri2024asped}, audio-based systems can detect the presence of pedestrians by using advances in acoustic scene analysis and deep learning, although challenges persist in signal separation, data imbalance, and generalizability across urban soundscapes.

The generalizability of pedestrian detection models has been explored only recently. Rasouli et al.\ assessed seven state-of-the-art detection algorithms under varying real-world conditions using the JAAD dataset and found that model performance deteriorates in changed contexts, such as different weather conditions, pedestrian behaviors, or occlusion \cite{rasouli2019}. They emphasized the importance of incorporating diverse training data, showing that general-purpose object detection models trained on broader datasets tend to generalize better than those trained narrowly on pedestrian-focused inputs. More recently, Hasan et al.\ conducted a cross-dataset evaluation of pedestrian detectors and similarly found that traditional models generalized poorly because their training source usually does not contain dense pedestrian volume \cite{hasan2021}. Interestingly, general-purpose object detectors, not trained for pedestrian detection, showed better cross-dataset performance, suggesting that varied training sources can improve model transferability. Although these studies do not focus on audio-based models, they emphasize that testing generalizability across datasets is crucial. In the context of audio-based sensing, this implication is particularly relevant, as urban soundscapes can vary considerably depending on the surrounding environment. 

\subsection{Audio-based Urban Sensing}

Urban sound has emerged as a rich source of information for understanding city life, complementing traditional visual or spatial data. Early urban noise studies primarily emphasized environmental health and policy, focusing on the quantification of noise pollution from road traffic, railways, and industrial sources \cite{Rulff2022, Hammer_Swinburn_Neitzel_2013, Jariwala_Syed_Pandya_Gajera_2021}. These works led to the development of standardized noise maps and public health guidelines (e.g.,\cite{WHO_2022}). However, beyond its value as a nuisance, urban sound is increasingly recognized as a medium that implies information about human activity, mobility patterns, and the social vibrancy of public spaces \cite{Radicchi2020, Aiello_Schifanella_Quercia_Aletta_2016}.

Recent advances in sensing technologies and machine learning enable granular, automated analysis of urban soundscapes. Projects such as SONYC (Sounds of New York City) \cite{bello2019} have established a baseline for classifying general urban sounds, using annotations for broad event categories that include speech-oriented human sounds. Extending this scope of urban audio analysis further, Han et al.\ and Seshadri et al.\ introduced audio-based methods for detecting pedestrian presence \cite{han2024audio, seshadri2024asped}. Their approach utilized a new large-scale dataset with pedestrian-focused annotations. This dataset is composed of continuous recordings from real-world walking environments, enable models to learn from the full range of implicit acoustic cues (both speech and non-speech) that signal pedestrian presence. Their results highlight the potential of microphone-based sensing as a low-cost, privacy-preserving, and scalable complement to camera-based systems.

Despite recent progress, the generalizability of models across diverse urban environments and the interpretability of these models remain underexplored. Understanding the level of generalizability and audio cues that trigger models to predict pedestrian presence is~---given the variety of urban soundscape---~crucial for building robust and interpretable systems. 

\section{DATASET}
\label{sec:dataset}

\begin{figure}
    \centering
    \includegraphics[width=\columnwidth]{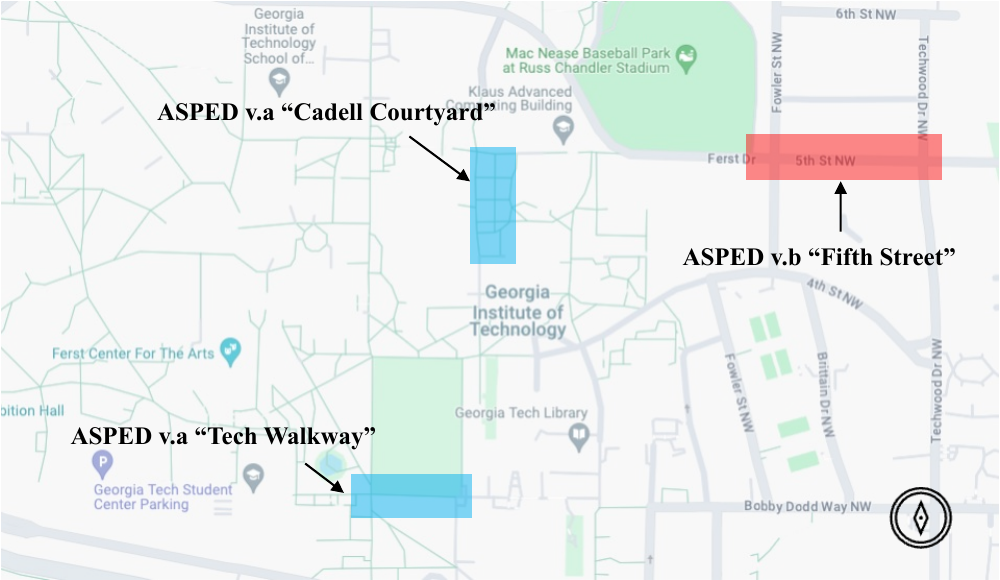} 
    \caption{Data collection sites on the Georgia Tech campus in Atlanta.}
    \label{fig:data_site}
\end{figure}

This study builds on the previously published ASPED dataset \cite{seshadri2024asped}, which includes annotated audio and video data collected in a vehicle-free courtyard environment and will be referred to in the following as ASPED v.a. This dataset provides the foundation for our pedestrian detection framework and is described in detail by Seshadri et al.~\cite{seshadri2024asped}. The recorder setup and preprocessing steps are identical to those used in ASPED v.a.

In this study, we introduce an additional dataset, ASPED v.b, collected close to a road with vehicular traffic. Figure~\ref{fig:data_site} highlights the recording location on the Georgia Tech campus in red. The vehicular noise primarily consists of engine sounds and intermittent shuttle buses operating at slow speeds. The proportion of frames containing at least one vehicle detected is 9.16\%, 29.00\%, 36.43\%, and 42.91\% for radii of \SI{1}{m}, \SI{3}{m}, \SI{6}{m}, and \SI{9}{m}, respectively.

The ASPED v.b dataset contains 1,321 hours of audio from 4 different sessions. Each session takes place over a time frame of approximately 40 hours and has audio data collected by 4 to 8 recorders spread along a street. The recording areas are monitored by 6 GoPro cameras, which captured \SI{1}{fps} video recordings totaling 2,946,513 frames across all cameras.

\begin{figure}
    \centering
    \includegraphics[width=1.1\columnwidth]{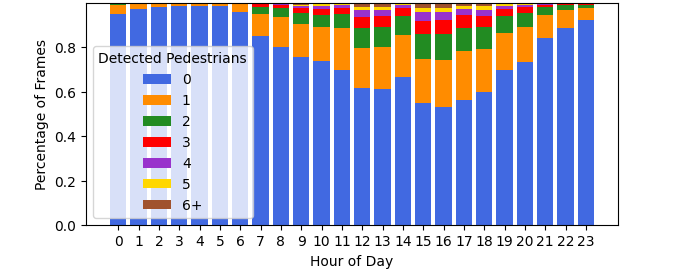} 
    \caption{Percentage of frames containing pedestrians by hour of day.}
    \label{fig:plot_hrs}
\end{figure}

Figure~\ref{fig:plot_hrs} illustrates general pedestrian patterns derived from the labels of the ASPED v.b dataset. The figure shows the ground truth number of pedestrians detected from video recordings at a specific timestamp, visualized for the recording zone with a \SI{6}{m} radius. Pedestrian activity peaks between 3 PM and 5 PM and declines considerably at night. This class imbalance reflects the ecological validity of the dataset, capturing realistic periods of low activity.

\begin{figure}
    \centering
    \includegraphics[width=\columnwidth]{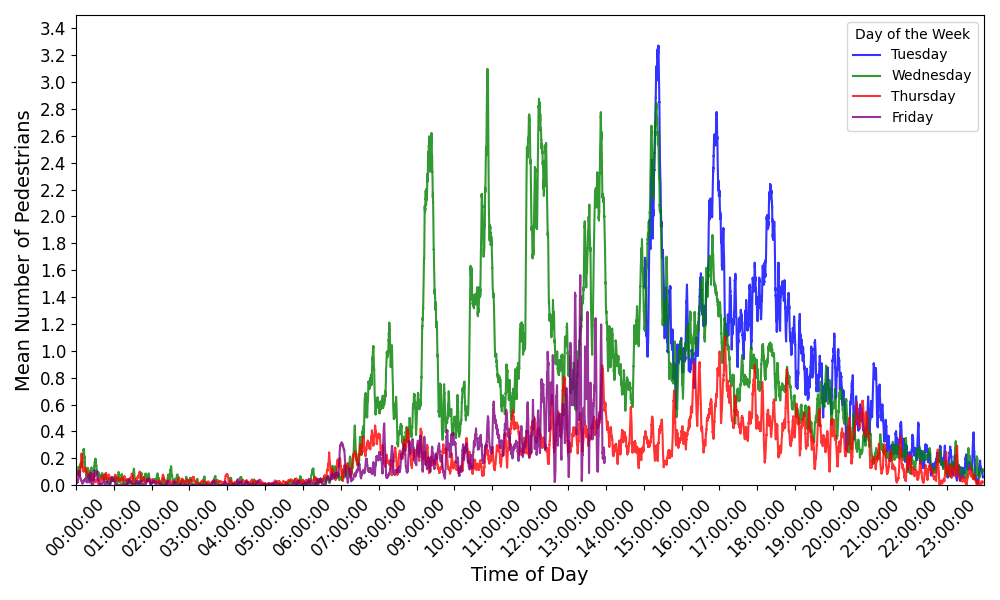} 
    \caption{Time-series distribution of pedestrian counts.}
    \label{fig:day_of_week_rolling_avg}
\end{figure}

The average number of pedestrians walking the street on a specific day of the week and time by taking the rolling average of the number of pedestrians detected across all cameras is shown in Fig.~\ref{fig:day_of_week_rolling_avg}. The peaks  align with the times that classes end on campus, demonstrating how pedestrian traffic on campus is closely tied to the class schedule.

Lastly, 2.9\% of total frames were obstructed by buses, preventing the video-based pedestrian annotation from producing reliable labels. Therefore, these frames were flagged and discarded in modeling.

\section{EXPERIMENTAL SETUP}
\label{sec:experimental_setup}

The goal of this research is to provide new insights into audio-based pedestrian detection that might facilitate new approaches with enhanced performance.
%Motivated by the creation of a dataset, this research aims to provide insights that can enhance performance in the pedestrian detection task. 
We conduct three key experiments to explore these aspects: 
\begin{inparaenum}[(i)]
\item a cross-dataset evaluation to assess the generalization capabilities of models trained on noisy and noise-free sections of the datasets (v.a and v.b), 
\item evaluating the effect of vehicle presence in training data on the performance with vehicle-controlled test sets, and 
\item an analysis of the acoustic cues that the model associates with pedestrian and non-pedestrian instances. 
\end{inparaenum}

For the experiments, we reproduced the model proposed by Seshadri et al.~\cite{seshadri2024asped}. This model processes 10-second \SI{16}{kHz} mono audio inputs by first computing power spectrograms using STFT (window: \SI{25}{ms}, hop: \SI{10}{ms}). These are then converted to 64-bin mel spectrograms (125--\SI{7500}{Hz}) and normalized via standard scaling. The mean and standard deviation values for normalization were obtained from the implementation provided in the GitHub repository of the ASPED v.a model.\footnote{\url{https://github.com/urbanaudiosensing/Models/blob/main/data_utils/transforms.py}, last access date: September 23, 2025} The resulting log-mel spectrograms are fed into the VGGish backbone, pre-trained on AudioSet \cite{gemmeke2017audioset}, to extract a sequence of 10 acoustic embeddings, each corresponding to a snippet of \SI{1}{s} of the input. A Transformer encoder (1 layer, 4 attention heads, 128 hidden dimension), with added positional encoding, processes these embeddings to capture temporal dependencies. Finally, a linear projection layer with ReLU activation, followed by another linear layer and a sigmoid activation function, outputs a binary classification probability for each \SI{1}{s} snippet, resulting in 10 predictions for one \SI{10}{s}-input, using a batch size of 256.

\subsection{Exp.~1: Cross-dataset evaluation}\label{subsec:cross-dataset-eval}
Following previously established methodology \cite{seshadri2024asped}, the two datasets, ASPED v.a and v.b, were randomly partitioned into train, test, and validation subsets with an 80/10/10 split, respectively. 

To address the inherent class imbalance, we employed weighted batch sampling and a variable weighted loss during training. Model inference results are reported using the checkpoint that yielded the lowest validation loss after 20 epochs.

\subsection{Exp.~2: Impact of vehicle presence for training} \label{subsec:impact_vehicle_presence}
A key difference between the previously existing dataset ASPED v.a and the new data lies in the presence of vehicle sounds in the audio recordings. In this experiment, we investigate whether this factor in the training data influences model performance on test environments with (VP: Vehicle-Present) and without vehicles (VA: Vehicle-Absent). To this end, we create two distinct test splits of ASPED v.b, controlled for vehicle presence and analyze the results for the models trained on v.a and v.b (cf.\ Sect.~\ref{subsec:cross-dataset-eval}), respectively. % were evaluated on these splits.

Furthermore, to assess the models' propensity for false positives, we sampled vehicle-related categories from the nonhuman sounds section of the FSD50K \cite{fonseca2022FSD50K} dataset. FSD50K  is an open dataset of human-labeled sound events containing 51,197 Freesound\footnote{\url{https://freesound.org/}, last access date: September 23, 2025} clips unequally distributed in 200 classes drawn from the AudioSet Ontology. For this and Section~\ref{subsec:what-model-hear}, we downsampled the audio to \SI{16}{kHz} and categorized it into human or non-human sounds by following the given ontology\footnote{\url{https://research.google.com/audioset/ontology/human_sounds_1.html}, last access date: September 23, 2025}. All classes in FSD50K are represented in AudioSet, except \textit{Crash cymbal} (non-human), \textit{Human group actions} (human), \textit{Human voice} (human), \textit{Respiratory sounds} (human), and \textit{Domestic sounds, home sounds} (non-human). Only single-tagged audio samples were included in this analysis, and we filtered the dataset to include only categories containing at least 10 distinct files. The resulting refined dataset comprised 21 human sound categories (989 files) and 133 non-human sound categories (8,097 files). The probability of class 1, which the model was trained to associate with `pedestrian' presence, was used to determine the model's response.

\begin{table}
\centering
\caption{Cross-dataset evaluation balanced accuracy (\%).}
\label{tab:cross_dataset}
\sisetup{
    reset-text-series = false,
    text-series-to-math = true,
    mode=text,
    tight-spacing=true,
    round-mode=places,
    round-precision=2,
    table-format=2.2,
    table-number-alignment=center
}
\begin{tabular}{l *{2}{S[table-column-width=2.85cm]}}
    \toprule
    \textbf{Train Dataset} & \multicolumn{2}{c}{\textbf{Test Dataset}} \\
    \cmidrule(l){2-3}
    & {\textbf{ASPED v.a $\uparrow$}} & {\textbf{ASPED v.b $\uparrow$}} \\
    \midrule
    \textbf{ASPED v.a} & 71.74 & 66.48 \\
    \textbf{ASPED v.b} & 64.77 & 69.15 \\
    \bottomrule
\end{tabular}
\end{table}

\subsection{Exp.~3: Model sensitivity to different sound categories}\label{subsec:what-model-hear}
To gain insights into ``what the models are listening to,'' we analyze the sensitivity of the ASPED-trained models to various audio categories by classifying inputs from the FSD50K human and non-human sound ontologies. More specifically, we investigate which human-generated sound categories were most frequently detected as `pedestrian.' Furthermore, we conduct a post-hoc analysis to determine if any non-human sound categories are consistently misclassified as `pedestrian.'

A crucial consideration for this analysis is the difference in both audio characteristics/recording setup and labeling paradigms. The ASPED dataset labels are based on the presence of individuals within a certain amount of radius of the recording device (in this study, \SI{6}{m}). In contrast, FSD50K annotations do not consider spatial proximity; for this evaluation, we operated under the assumption that all human sounds represent the `pedestrian' class and all non-human sounds represent the `non-pedestrian' class.

We further investigate the specific categories of human-related sounds that our model reliably detects or struggles to recognize. Additionally, we examine non-human sounds that are erroneously classified as pedestrian-related, leading to false positive errors.

\section{RESULTS}
\label{sec:results}

\subsection{Exp.~1: Cross-dataset evaluation}\label{subsec:result-cross-dataset-eval}

Table~\ref{tab:cross_dataset} presents the balanced accuracy, calculated as the average of sensitivity and specificity, achieved when models trained on one dataset version were tested on the other.

The results indicate a performance drop when models are tested on a dataset different from their training set, which indicates limited generalization across the two recording setups.

These cross-dataset results highlight the complex interplay between the presence of specific types of background noise, such as vehicular traffic, and model generalization. Further investigation into domain adaptation techniques may be beneficial to improve the robustness of pedestrian detection systems in real-world scenarios with varying acoustic environments.

\subsection{Exp.~2: Impact of vehicle presence for training} \label{subsec:result-impact_vehicle_presence}

To investigate the specific impact of vehicle presence in the training data, we evaluate the v.a-trained and v.b-trained models on subsets of v.b that were controlled for the presence or absence of vehicle sounds (VP: Vehicle-Present, VA: Vehicle-Absent), as shown in Table~\ref{tab:impact_vehicle_presence_results}.

\begin{table}
\centering
\caption{Impact of vehicle presence in training data --- balanced accuracy (\%) on ASPED v.b subsets. (VP: Vehicle-Present, VA: Vehicle-Absent)}
\label{tab:impact_vehicle_presence_results}
\sisetup{
    reset-text-series = false,
    text-series-to-math = true,
    mode=text,
    tight-spacing=true,
    round-mode=places,
    round-precision=2,
    table-format=2.2,
    table-number-alignment=center
}
\begin{tabular}{l *{2}{S[table-column-width=2.85cm]}}
    \toprule
    \textbf{Train Dataset} & \multicolumn{2}{c}{\textbf{Test Dataset (ASPED v.b)}} \\
    \cmidrule(l){2-3}
    & {\textbf{VP $\uparrow$}} & {\textbf{VA $\uparrow$}} \\
    \midrule
    \textbf{ASPED v.a} & 65.16 & 67.87 \\
    \textbf{ASPED v.b} & 67.49 & 71.01 \\
    \bottomrule
\end{tabular}
\end{table}

The results show that~---as expected---~the presence or absence of vehicle sounds in the test set impacts performance. Even though the v.b-trained model was exposed to vehicle sounds during training, predicting pedestrian presence in the absence of these potentially confounding sounds is simpler.

To assess whether the v.b-trained model exhibits a reduced tendency to misclassify common vehicle sounds as pedestrians compared to the v.a-trained model, we compared the average predicted probability of the `pedestrian' class for a curated set of vehicle-related non-human sound categories from FSD50K (Table~\ref{tab:audioset_vehicle_comparison}).

\begin{table}
\centering
\caption{Avg.\ prob. of pedestrian class for vehicle-related FSD50K categories.}
\label{tab:audioset_vehicle_comparison}
\sisetup{
    reset-text-series = false,
    text-series-to-math = true,
    mode=text,
    tight-spacing=true,
    round-mode=places,
    round-precision=2,
    table-format=1.2,
    table-number-alignment=center
}
\begin{tabular*}{\columnwidth}{l @{\extracolsep{\fill}}cc}
%\begin{tabular}{l S S}
    \toprule
    \textbf{Category} & {\textbf{v.a-trained $\downarrow$}} & {\textbf{v.b-trained $\downarrow$}} \\
    \midrule
    \textit{Race car, auto racing} & 0.86 & 0.69 \\
    \textit{Car} & 0.69 & 0.62 \\
    \textit{Vehicle} & 0.75 & 0.56 \\
    \textit{Vehicle horn, car horn, honking} & 0.74 & 0.62 \\
    \textit{Car passing by} & 0.71 & 0.59 \\
    \textit{Motor vehicle (road)} & 0.72 & 0.60 \\
    \bottomrule
\end{tabular*}
\end{table}

The v.a-trained model generally exhibited considerably higher average predicted probabilities for classifying vehicle sounds as `pedestrian' compared to the v.b-trained model. This suggests that the absence of traffic noise during training in v.a might lead the model to erroneously associate vehicle sounds with human presence, increasing false positives. Conversely, the v.b-trained model trained with traffic noise was more effective at distinguishing pedestrian presence from vehicle sounds as indicated by fewer false alarms.

\subsection{Exp.~3: Model sensitivity to different sound categories}\label{subsec:result-what-model-hear}
% The models were evaluated on an out-of-domain dataset, split into human and non-human sounds, to identify the impact of different sound categories on the likelihood of detecting pedestrian presence.
To understand the models' sensitivity to different acoustic cues, we investigated the impact of signal energy and of different (human and non-human) sounds on the pedestrian detection accuracy.

\subsubsection{Comparison with RMS energy}
{The Pearson correlation between the audio's RMS energy and the model's output logit is low for models trained on ASPED v.a and v.b ($r \approx 0.14$ and $r \approx 0.29$, respectively), confirming that the learned representations are more effective than a simple energy measurement. }

\subsubsection{Evaluation on FSD50K Human Sounds}
Table~\ref{tab:audioset_avg_prob_human} presents the human categories as a subset of the FSD50K dataset. On the right, we list the corresponding average predicted probability of the `pedestrian' class for both models.
\begin{table}
\centering
\caption{Average probability for FSD50K human sound categories for models trained on ASPED v.a and v.b.}
\label{tab:audioset_avg_prob_human}
\sisetup{
    reset-text-series = false,
    text-series-to-math = true,
    mode=text,
    tight-spacing=true,
    round-mode=places,
    round-precision=2,
    table-format=1.2,
    table-number-alignment=center
}
\begin{tabular*}{\columnwidth}{l @{\extracolsep{\fill}}cc}
%\begin{tabular}{l S S}
    \toprule
    \textbf{Category} & {\textbf{v.a-trained  $\uparrow$}} & {\textbf{v.b-trained  $\uparrow$}} \\
    \midrule
    \textit{Female singing} & 0.95 & 0.74 \\
    \textit{Speech} & 0.94 & 0.65 \\
    \textit{Crying, sobbing} & 0.92 & 0.63 \\
    \textit{Laughter} & 0.91 & 0.65 \\
    \textit{Singing} & 0.90 & 0.71 \\
    \textit{Human voice} & 0.89 & 0.64 \\
    \textit{Yell} & 0.89 & 0.66 \\
    \textit{Cheering} & 0.88 & 0.66 \\
    \textit{Chatter} & 0.87 & 0.66 \\
    \textit{Child speech, kid speaking} & 0.86 & 0.65 \\
    \textit{Human group actions} & 0.83 & 0.63 \\
    \textit{Speech synthesizer} & 0.82 & 0.59 \\
    \textit{Conversation} & 0.79 & 0.63 \\
    \textit{Burping, eructation} & 0.78 & 0.56 \\
    \textit{Male speech, man speaking} & 0.77 & 0.56 \\
    \textit{Whispering} & 0.76 & 0.52 \\
    \textit{Applause} & 0.76 & 0.48 \\
    \textit{Chewing, mastication} & 0.76 & 0.57 \\
    \textit{Hands} & 0.74 & 0.56 \\
    \textit{Run} & 0.74 & 0.56 \\
    \textit{Walk, footsteps} & 0.70 & 0.58 \\
    \bottomrule
\end{tabular*}
\end{table}
The model trained on the ASPED v.a dataset demonstrates greater confidence when classifying human sounds as `pedestrian' compared to its counterpart trained on the traffic-noise-rich ASPED v.b dataset.
While speech-related sounds generally exhibited higher probabilities across both models, subtle performance variations in the ranking of specific categories might indicate that background noise during training influences the model's sensitivity to different types of human sounds. The overall lower average probabilities for the v.b-trained model likely reflect the masking effect of traffic noise on the acoustic features crucial for human sound identification.
Notably, categories intuitively associated with pedestrian movement, such as \textit{Walk, footsteps} and \textit{Run}, were ranked relatively low within the broader set of human sound categories for both models.
These findings underscore the impact of the training environment's acoustic characteristics on the learned representations and the subsequent generalization to out-of-domain human sounds. It should be noted, however, that the majority of signals in this dataset are very different from the typical urban sound recording; thus, these results should be interpreted carefully.

%\vspace{0.5em}
\subsubsection{Evaluation on FSD50K Non-Human Sounds}
To understand the models' sensitivity to other sounds, we evaluated their predictions on a subset of 133 (categories with at least 10 samples) non-human sound categories from AudioSet. Table~\ref{tab:audioset_top_bottom_nonhuman_va_vb} displays the top 3 and bottom 3 categories, determined based on the v.a-trained model's average predicted probability of the `pedestrian' class.
\begin{table}
\centering
\caption{Top and bottom 3 non-human sound categories by avg. prob.}
\label{tab:audioset_top_bottom_nonhuman_va_vb}
\begin{tabular*}{\columnwidth}{l @{\extracolsep{\fill}}cc}
%\begin{tabular}{l S S}
    \toprule
    \textbf{Category} & {\textbf{v.a-trained  $\downarrow$}} & {\textbf{v.b-trained  $\downarrow$}} \\
    \midrule
    \multicolumn{3}{l}{\textbf{Top 3}} \\
    \textit{Harp} & 0.94 $\pm$ 0.07 & 0.71 $\pm$ 0.13 \\
    \textit{Trumpet} & 0.94 $\pm$ 0.14 & 0.80 $\pm$ 0.11 \\
    \textit{Plucked string instrument} & 0.93 $\pm$ 0.07 & 0.66 $\pm$ 0.15 \\
    \midrule
    \multicolumn{3}{l}{\textbf{Bottom 3}} \\
    \textit{Cricket} & 0.42 $\pm$ 0.33 & 0.48 $\pm$ 0.16 \\
    \textit{Chirp, tweet} & 0.51 $\pm$ 0.31 & 0.45 $\pm$ 0.15 \\
    \textit{Bicycle bell} & 0.52 $\pm$ 0.36 & 0.48 $\pm$ 0.14 \\
    \bottomrule
\end{tabular*}
\end{table}
The evaluation on non-human sounds reveals that the model trained on v.a data has a higher tendency to misclassify certain musical instruments as `pedestrian' compared to the v.b-trained model. This may be due to such sound categories being particularly infrequent or entirely absent in the ASPED datasets.
Interestingly, the bottom-ranked categories reveal greater prediction variability in the v.a-trained model compared to the v.b-trained model. This higher standard deviation suggests that the v.a model is less certain when classifying sounds that are dissimilar to human presence.

\section{CONCLUSION}
\label{sec:conclusion}

This research investigated the impact of the acoustic environment on pedestrian detection using a novel pedestrian detection dataset with vehicular noise. Our cross-dataset evaluation revealed a performance drop when models were trained on different environments, indicating limited domain generalization capability. Furthermore, the presence of vehicle sounds in the test set considerably influenced performance, with models showing varying sensitivities based on their training data's acoustic characteristics. Evaluation on out-of-domain FSD50K data highlighted that models trained in v.a exhibited higher confidence in identifying human sounds but were also more prone to false positives for non-pedestrian sounds. Conversely, models trained with traffic noise demonstrated more cautious predictions. However, the notable issue of false positives across various non-human sound categories warrants further attention. These findings underscore the critical role of the acoustic environment in training robust pedestrian detection systems. The limited generalization observed suggests that future work should focus on domain adaptation techniques to bridge the gap between different acoustic domains. Specifically, exploring methods to enhance the model's ability to filter out irrelevant background noise, such as vehicular traffic, while retaining sensitivity to subtle pedestrian-related cues is crucial. Additionally, we plan to investigate the integration of multi-modal information (e.g., visual cues) to increase robustness in challenging scenarios. Finally, a more comprehensive analysis of the model's failure cases, particularly the misclassification of specific non-human sounds, could inform the design of more discriminative acoustic features or robust model architectures.

% \section{Acknowledgment}
% \label{sec:ack}

% The preferred spelling of the word acknowledgment in America is without an ``e'' after the ``g.'' Try to avoid the stilted expression, ``One of us (R.\ B.\ G.) thanks ...'' Instead, try ``R.\ B.\ G.\ thanks ...''  Put sponsor acknowledgments in the unnumbered footnote on the first page. Please include acknowledgments only in the camera-ready version, and NOT in the version of the paper submitted for review.

% -------------------------------------------------------------------------
% Either list references using the bibliography style file IEEEtran.bst

\clearpage
% The \IEEEtriggeratref{XX} command can be used to move to the next column before the XX-th reference
% to balance the two columns of the reference section
% \IEEEtriggeratref{XX}
\bibliographystyle{IEEEtran}
\bibliography{refs}

\begin{thebibliography}{10}
\providecommand{\url}[1]{#1}
\csname url@samestyle\endcsname
\providecommand{\newblock}{\relax}
\providecommand{\bibinfo}[2]{#2}
\providecommand{\BIBentrySTDinterwordspacing}{\spaceskip=0pt\relax}
\providecommand{\BIBentryALTinterwordstretchfactor}{4}
\providecommand{\BIBentryALTinterwordspacing}{\spaceskip=\fontdimen2\font plus
\BIBentryALTinterwordstretchfactor\fontdimen3\font minus \fontdimen4\font\relax}
\providecommand{\BIBforeignlanguage}[2]{{%
\expandafter\ifx\csname l@#1\endcsname\relax
\typeout{** WARNING: IEEEtran.bst: No hyphenation pattern has been}%
\typeout{** loaded for the language `#1'. Using the pattern for}%
\typeout{** the default language instead.}%
\else
\language=\csname l@#1\endcsname
\fi
#2}}
\providecommand{\BIBdecl}{\relax}
\BIBdecl

\bibitem{apa1965pedestriancount}
{American Planning Association}, ``The pedestrian count,'' \url{https://www.planning.org/pas/reports/report199.htm}, 1965.

\bibitem{gdpr2019guidelines}
\BIBentryALTinterwordspacing
{GDPR.eu}, ``General data protection regulation (gdpr) compliance guidelines,'' 2019, last access date: May 7, 2025. [Online]. Available: \url{https://gdpr.eu/}
\BIBentrySTDinterwordspacing

\bibitem{li2016pedestrian}
H.~Li, Z.~Wu, and J.~Zhang, ``Pedestrian detection based on deep learning model,'' in \emph{International Congress on Image and Signal Processing, BioMedical Engineering and Informatics (CISP-BMEI)}.\hskip 1em plus 0.5em minus 0.4em\relax IEEE, 2016, pp. 796--800.

\bibitem{seshadri2024asped}
P.~Seshadri, C.~Han, B.-W. Koo, N.~Posner, S.~Guhathakurta, and A.~Lerch, ``Asped: An audio dataset for detecting pedestrians,'' in \emph{International Conference on Acoustics, Speech and Signal Processing (ICASSP)}, 2024, pp. 406--410.

\bibitem{yang2010automatic}
H.~Yang, K.~Ozbay, and B.~Bartin, ``Investigating the performance of automatic counting sensors for pedestrian traffic data collection,'' in \emph{World Conference on Transport Research (WCTR)}, vol. 1115, 2010, pp. 1--11.

\bibitem{yang2011infrared}
------, ``Enhancing the quality of infrared-based automatic pedestrian sensor data by nonparametric statistical method,'' \emph{Transportation Research Record: Journal of the Transportation Research Board}, vol. 2264, no.~1, pp. 11--17, 2011.

\bibitem{BRUNETTI201817}
A.~Brunetti, D.~Buongiorno, G.~F. Trotta, and V.~Bevilacqua, ``Computer vision and deep learning techniques for pedestrian detection and tracking: A survey,'' \emph{Neurocomputing}, vol. 300, pp. 17--33, 2018.

\bibitem{5975165}
P.~Dollar, C.~Wojek, B.~Schiele, and P.~Perona, ``Pedestrian detection: An evaluation of the state of the art,'' \emph{IEEE Transactions on Pattern Analysis and Machine Intelligence (TPAMI)}, vol.~34, no.~4, pp. 743--761, 2011.

\bibitem{ozan2021}
E.~Ozan, S.~Searcy, B.~C. Geiger, C.~Vaughan, C.~Carnes, C.~Baird, and A.~Hipp, ``State-of-the-art approaches to bicycle and pedestrian counters,'' North Carolina Department of Transportation, Tech. Rep., 2021.

\bibitem{rasouli2019}
A.~Rasouli, I.~Kotseruba, and J.~K. Tsotsos, ``It's not all about size: On the role of data properties in pedestrian detection,'' in \emph{European Conference on Computer Vision (ECCV) Workshops}, 2018, pp. 210--225.

\bibitem{hasan2021}
I.~Hasan, S.~Liao, J.~Li, S.~U. Akram, and L.~Shao, ``Generalizable pedestrian detection: The elephant in the room,'' in \emph{Conference on Computer Vision and Pattern Recognition (CVPR)}, 2021, pp. 11\,323--11\,332.

\bibitem{Rulff2022}
J.~Rulff, F.~Miranda, M.~Hosseini, M.~Lage, M.~Cartwright, G.~Dove, J.~Bello, and C.~T. Silva, ``Urban rhapsody: Large-scale exploration of urban soundscapes,'' \emph{Computer Graphics Forum}, vol.~41, no.~3, pp. 209--221, 2022.

\bibitem{Hammer_Swinburn_Neitzel_2013}
M.~S. Hammer, T.~K. Swinburn, and R.~L. Neitzel, ``Environmental noise pollution in the united states: developing an effective public health response,'' \emph{Environmental Health Perspectives (EHP)}, vol. 122, no.~2, pp. 115--119, 2014.

\bibitem{Jariwala_Syed_Pandya_Gajera_2021}
\BIBentryALTinterwordspacing
H.~J. Jariwala, H.~S. Syed, M.~J. Pandya, and Y.~M. Gajera, \emph{Noise Pollution \& Human Health: A review}, 2021, last access date: May 7, 2025. [Online]. Available: \url{https://www.researchgate.net/profile/Hiral-Jariwala/publication/319329633_Noise_Pollution_Human_Health_A_Review/links/59a54434a6fdcc773a3b1c49/Noise-Pollution-Human-Health-A-Review.pdf}
\BIBentrySTDinterwordspacing

\bibitem{WHO_2022}
\BIBentryALTinterwordspacing
{World Health Organization}, ``Environmental noise,'' in \emph{Compendium of {WHO} and other {UN} guidance on health and environment}, 2022, ch.~11, last access date: May 7, 2025. [Online]. Available: \url{https://cdn.who.int/media/docs/default-source/who-compendium-on-health-and-environment/who_compendium_noise_01042022.pdf?sfvrsn=bc371498_3#:~:text=For%20average%20noise%20exposure%2C%20the,dB%20LAeq%2C%2024h%20%E2%80%A2%20weekly}
\BIBentrySTDinterwordspacing

\bibitem{Radicchi2020}
A.~Radicchi, P.~Cevikayak~Yelmi, A.~Chung, P.~Jordan, S.~Stewart, A.~Tsaligopoulos, L.~McCunn, and M.~Grant, ``Sound and the healthy city,'' \emph{Cities \& Health}, vol.~5, no. 1-2, pp. 1--13, 2021.

\bibitem{Aiello_Schifanella_Quercia_Aletta_2016}
L.~M. Aiello, R.~Schifanella, D.~Quercia, and F.~Aletta, ``Chatty maps: constructing sound maps of urban areas from social media data,'' \emph{Royal Society Open Science}, vol.~3, no.~3, p. 150690, 2016.

\bibitem{bello2019}
J.~P. Bello, C.~Silva, O.~Nov, R.~L. Dubois, A.~Arora, J.~Salamon, C.~Mydlarz, and H.~Doraiswamy, ``Sonyc: A system for monitoring, analyzing, and mitigating urban noise pollution,'' \emph{Communications of the ACM}, vol.~62, no.~2, pp. 68--77, 2019.

\bibitem{han2024audio}
C.~Han, P.~Seshadri, Y.~Ding, N.~Posner, B.~W. Koo, A.~Agrawal, A.~Lerch, and S.~Guhathakurta, ``Understanding pedestrian movement using urban sensing technologies: the promise of audio-based sensors,'' \emph{Urban Informatics}, vol.~3, no.~1, p.~22, 2024.

\bibitem{gemmeke2017audioset}
J.~F. Gemmeke, D.~P. Ellis, D.~Freedman, A.~Jansen, W.~Lawrence, R.~C. Moore, M.~Plakal, and M.~Ritter, ``Audio set: An ontology and human-labeled dataset for audio events,'' in \emph{International Conference on Acoustics, Speech and Signal Processing (ICASSP)}, 2017, pp. 776--780.

\bibitem{fonseca2022FSD50K}
E.~Fonseca, X.~Favory, J.~Pons, F.~Font, and X.~Serra, ``Fsd50k: an open dataset of human-labeled sound events,'' \emph{IEEE/ACM Transactions on Audio, Speech, and Language Processing (TASLP)}, vol.~30, pp. 829--852, 2022.

\end{thebibliography}
% or list them by yourself:

% \bibitem{dcaseweb}
% {DCASE Website}, \url{http://www.dcase.com}.

% \bibitem{IEEEXploreReqs}
% {IEEE {X}plore {R}equirements}, \url{https://conferences.ieeeauthorcenter.ieee.org/write-your-paper/meet-ieee-xplore-requirements/}.

% \bibitem{eWilliams1999}
% E.~Williams, \emph{Fourier Acoustics: Sound Radiation and Nearfield Acoustic Holography}.\hskip 1em plus 0.5em minus 0.4em\relax London, UK: Academic Press, 1999.

% \bibitem{cJones2003}
% C.~Jones, A.~Smith, and E.~Roberts, ``A sample paper in conference proceedings,'' in \emph{Proc. ICASSP}, vol.~II, Apr. 2003, pp. 803--806.

% \bibitem{aSmith2000}
% A.~Smith, C.~Jones, and E.~Roberts, ``A sample paper in journals,'' \emph{IEEE Trans. Signal Process.}, vol.~62, pp. 291--294, Jan. 2000.

% \end{thebibliography}

\end{document}